\documentclass[12pt,preprint]{aastex}

\newcommand{\sh}{\mbox{Sh\,2-216}}

\slugcomment{Accepted for publication in the Astrophysical Journal.}

\shorttitle{Probing the ISM around the PN \sh}
\shortauthors{Ransom et al.}

\begin{document}
      
\title{PROBING THE MAGNETIZED INTERSTELLAR MEDIUM SURROUNDING THE
  PLANETARY NEBULA SH\,2-216}

\author{R. R. Ransom\altaffilmark{1}, B. Uyan{\i}ker\altaffilmark{1,2},
  R. Kothes\altaffilmark{1,3} and T. L. Landecker\altaffilmark{1}}

\altaffiltext{1}{National Research Council of Canada, Herzberg Institute
  of Astrophysics, Dominion Radio Astrophysical Observatory, Box 248,
  Penticton, BC, V2A 6J9, Canada}

\altaffiltext{2}{Present Address: 35-3737 Gellatly Road, Westbank, BC,
  V2T 2W8, Canada}

\altaffiltext{3}{Department of Physics and Astronomy, University of
  Calgary, 2500 University Drive NW, Calgary, AB, T2N 1N4, Canada}

\email{Ryan.Ransom@nrc-cnrc.gc.ca}

\begin{abstract}
  
  We present 1420~MHz polarization images of a
  $2.5\arcdeg\times2.5\arcdeg$ region around the planetary nebula (PN)
  \sh.  The images are taken from the Canadian Galactic Plane Survey
  (CGPS).  An arc of low polarized intensity (size $0.2\arcdeg \times
  0.7\arcdeg$) appears prominently in the north-east portion of the
  visible disk of \sh, coincident with the optically identified
  interaction region between the PN and the interstellar medium (ISM).
  The arc contains structural variations down to the $\sim$1\arcmin\@
  resolution limit in both polarized intensity and polarization angle.
  Several polarization-angle ``knots'' appear along the arc.  By
  comparison of the polarization angles at the centers of the knots
  and the mean polarization angle outside \sh, we estimate the $RM$
  through the knots to be $-43 \pm 10\ \rm{rad}\,\rm{m}^{-2}$.  Using
  this estimate for the $RM$ and an estimate of the electron density
  in the shell of \sh, we derive a line-of-sight magnetic field in the
  interaction region of $5.0 \pm 2.0$~$\mu\rm{G}$.  We believe it more
  likely the observed magnetic field is interstellar than stellar,
  though we cannot completely dismiss the latter possibility.  We
  interpret our observations via a simple model which describes the
  ISM magnetic field around \sh, and comment on the potential use of
  old PNe as probes of the magnetized ISM.

\end{abstract}


\section{INTRODUCTION\label{intro}}

The diffuse Galactic synchrotron radiation provides a continuous
background of radio emission which is intrinsically highly (up to
$\approx$70\%) linearly polarized.  This radiation is Faraday-rotated
from the point of emission as it propagates through warm ionized gas
interwoven with magnetic fields in the disk of the Galaxy; i.e., the
angle $\theta$ of the polarized component of the emission is rotated
at wavelength $\lambda$\,[m] by
\begin{equation}
\Delta\theta = RM\,\lambda^2\ [\rm{rad}],
\end{equation}
where $RM$ is the rotation measure [$\rm{rad}\,\rm{m}^{-2}$] and
depends on the line-of-sight component of the magnetic field,
$B_{\|}$\,$[\mu\rm{G}]$, the thermal electron density,
$n_e$\,$[\rm{cm}^{-3}]$, and the path length, $dl$\,$[\rm{pc}]$, as
\begin{equation}
RM = 0.81\int{B_{\|}\,n_e\,dl\ [\rm{rad}\,\rm{m}^{-2}]}.
\end{equation}
High-resolution radio polarization images at frequencies
$\lesssim$3~GHz reveal the turbulent imprint of Faraday rotation on
the diffuse polarized emission (e.g., \citealt{Wieringa+1993};
\citealt{Gray+1999}; \citealt{Gaensler+2001}; \citealt{Uyaniker+2003};
\citealt*{HaverkornKd2003a,HaverkornKd2003c};
\citealt{Haverkorn+2006b}; \citealt{Schnitzeler+2007}).  The turbulent
nature of the imprint is the product of the random component of the
Galactic magnetic field and irregular electron-density distributions
in the general interstellar medium (ISM).  Detailed studies and
modeling of the diffuse Galactic emission (e.g.,
\citealt{Spoelstra1984}; \citealt*{HaverkornKd2004b}) as well as
statistical analyses of the $RM$s of polarized extragalactic sources
\citep{Haverkorn+2006a} indicate that the scale size, or ``cell''
size, for variations in the magnetized ISM range from $\sim$15~pc to
100~pc.  Depth depolarization then results from the averaging of
nonparallel polarization vectors from emission at different cells
along the line-of-sight.  Smaller-scale variations are also apparent
in polarization images.  Depolarization filaments, or ``canals,'' with
a width corresponding to one beam of the observing instrument,
indicate the presence of very sharp gradients in $RM$ (see, e.g.,
\citealt{Gaensler+2001}; \citealt{Uyaniker+2003};
\citealt*{HaverkornKd2004a}).  If the gradient is so steep across the
beam as to cause differential rotation of the polarization angle of
$\sim$90\arcdeg, then complete depolarization occurs.  Beam
depolarization in images produced by aperture synthesis telescopes
(typically with arcminute resolution) suggests a scale length for $RM$
structures in the magnetized ISM of less than 1~pc.

Interpreting the structure seen in radio polarization images of the
Galactic plane is largely left to modeling
\citep[e.g.,][]{HaverkornKd2004b}, as there is generally little
correlation between the polarization structures seen in these images
and the emission structures seen in total intensity images.
Nevertheless, objects of known distance can be used to estimate the
line-of-sight distribution of the magnetized ISM, revealing, for
example, whether observed polarization structures are generated behind
the known object, in the region between the object and the Sun, or
perhaps in the immediate vicinity of the object
\citep[see][]{Gray+1999,UyanikerL2002b}.  Moreover, if small-scale
depolarization structures can be isolated to a known object, then it
may be possible to probe directly the properties of the magnetized ISM
within the structures.

\ion{H}{2} regions and supernova remnants (SNRs) are the two most
prevalent (discrete) constituents of the ISM as seen at radio
wavelengths, and each class of object is detected in radio
polarization images up to a limiting distance determined by the
``polarization horizon'' \citep*[see][]{KothesL2004}. \ion{H}{2}
regions are detected by way of their depolarizing effects on
background diffuse emission, while SNRs are simultaneously a source of
polarized synchrotron emission and a Faraday ``screen'' which
depolarizes background emission.  However, neither class of object is
a particularly good probe of the magnetized ISM.  \ion{H}{2} regions
have very high electron densities and turbulent motions which produce
tangled magnetic fields, a combination which leads to virtually
complete beam depolarization across the region.  For SNRs, complex
models are needed to describe the physical parameters in the shock
front at the interface between the rapidly expanding SNR and the ISM.

Another class of object which may potentially serve as a better probe
of the magnetized ISM is planetary nebulae (PNe).  Young PNe are
relatively strong thermal radio emitters, resulting from high electron
densities \citep[see, e.g.,][]{Bains+2003}, but they are also very
compact ($\ll$1~pc) and not yet interacting with the ISM.  On the
other hand, the shells of many old PNe are observed at optical
wavelengths to interact with the ISM \citep[see][]{TweedyK1996}.
Moreover, the ISM magnetic field appears to play a significant role in
shaping the shells, as evidenced primarily by the visible ``striping''
or filamentary structure of shell gases
\citep*[see][]{TweedyMN1995,SokerZ1997}.  Theoretical treatments show
that interactions between PNe and the ISM are an important
consideration in the evolution of PN systems moving at even modest
speeds ($\gtrsim$5~$\rm{km}\,\rm{s}^{-1}$) with respect to the ISM
\citep*[see, e.g.,][]{SokerD1997,WareingZO2007}.  If the conditions
are right in the interaction region between the PN and the ISM, we may
expect to see the Faraday signature of old PNe in radio polarization
images.  Such a signature has been identified for the nearby PN
Sharpless~2-216 (\sh) and was first described by one of us in
\citet{Uyaniker2004}.  In this paper, we describe in detail the
Faraday-rotation structure in the shell of \sh.

We present radio polarization images at 1420~MHz of a
$2.5\arcdeg\times2.5\arcdeg$ region of the Galactic plane around the
position of \sh.  The images are taken from the Canadian Galactic
Plane Survey (CGPS).  In \S~\ref{s216}, we summarize the pertinent
properties of \sh.  In \S~\ref{obs}, we describe briefly the
preparation of the images.  In \S~\ref{results}, we describe the
structures observed on the visible disk of \sh\@ in both polarized
intensity and polarization angle, and estimate the $RM$ through the
shell of \sh.  In \S~\ref{discuss}, we derive the magnetic field in
the shell of \sh, and interpret the structures and $RM$s in the
context of an interaction between the PN and the ISM.  We also discuss
the possibility that the observed structures are produced by the
stellar field of the host white dwarf or its progenitor.  Finally, in
\S~\ref{concl}, we summarize our conclusions.

\section{THE PLANETARY NEBULA SH\,2-216 \label{s216}}

\sh\@ is the closest known PN.  At a distance of 129~pc
\citep{Harris+2007}, its 1.7$\arcdeg$ angular diameter translates to a
physical diameter of 3.8~pc, making it also one of the largest and
oldest PNe.  The most conspicuous feature of \sh\@ at optical
wavelengths is its bright eastern rim, denoting an interaction between
the expanding and moving PN and the ISM.  The location of the
interaction region\footnote{We refer to the bright eastern rim as
  ``the interaction region'' throughout the paper, though other parts
  of the shell of \sh\@ may also be interacting with the ISM.}
appears to be consistent with the observed displacement of the host
white dwarf from the center of the PN; i.e., the enhanced emission in
this region is a consequence of the additive velocity of the nebular
expansion in all directions and the underlying eastward motion of the
PN system relative to the ISM.  The expansion velocity is very low
\citep[$<4~\rm{km}\,\rm{s}^{-1}$;][]{Reynolds1985}, indicating that
the ISM pressure (with dynamic, magnetic and cosmic-ray components) is
nearly equal to the PN ram pressure.  The velocity (on the plane of
the sky) of the system relative to the ISM is estimated also to be
$\sim$4~$\rm{km}\,\rm{s}^{-1}$ \citep{TweedyMN1995}.

The thin filamentary structures observed in $\rm{H}\alpha$ in the
interaction region, together with the more subtle, and wider,
filamentary structure observed in \ion{N}{2} across the face of the
PN, qualitatively suggest that the ISM magnetic field is shaping the
morphology of \sh\@ \citep{TweedyMN1995}.  Using estimates for the
electron density within the PN ($n_e \sim 5~\rm{cm}^{-3}$) and the
mean ISM magnetic field ($B \sim 5~\mu\rm{G}$), \citet{TweedyMN1995}
show that the ISM magnetic pressure is about twice that of the dynamic
pressure, and thus likely a dominant factor in the shaping.  The
strength of the PN magnetic field is expected to be negligible at
large radii ($r \sim 1$~pc) from the host white dwarf, assuming the
field decreases as $B \sim r^{-2}$ \citep*[see,
e.g.,][]{VlemmingsDL2002}, but may be amplified significantly within
filaments due to compression \citep[see][]{Soker2002,HugginsM2005}.
We discuss the strength and orientation of the magnetic field in the
outermost regions of \sh\@ in \S~\ref{discuss}.

\section{OBSERVATIONS AND IMAGE PREPARATION \label{obs}}

The radio polarization data presented in this paper were obtained at
1420~MHz ($\lambda = 21$\,cm) as part of the CGPS \citep{Taylor+2003}
using the synthesis telescope (ST) at the Dominion Radio Astrophysical
Observatory (DRAO).  The ST is described in detail by
\citet{Landecker+2000}.  Images are produced in each CGPS field for
the two orthogonal linear polarization states, Stokes-Q ($Q$) and
Stokes-U ($U$), as well as Stokes-I (total intensity), from data in
each of four 7.5~MHz continuum bands centered on 1406.65, 1414.15,
1426.65 and 1434.15~MHz, respectively\footnote{Note that the frequency
  corresponding to the midpoint of the four continuum bands is
  1420.4~MHz, the neutral hydrogen spin-flip frequency.  The 5.0~MHz
  band about this frequency is allocated to the 256-channel
  spectrometer \citep[see][]{Taylor+2003}}.  Images in Stokes-V,
nominally representing circular polarization, are presently dominated
by instrumental errors, and are of significantly lower value.  The
DRAO ST is sensitive at 1420~MHz to emission from structures with
angular sizes of $\sim$1\arcdeg\@ (corresponding to the shortest,
12.9~m, baseline of the ST) down to the resolution limit of
$\sim$1\arcmin\@ (corresponding to the longest, 617.1~m, baseline).
In total intensity, data from the Effelsberg 21-cm Radio Continuum
Survey \citep{ReichRF1997} are added to the band-averaged ST data to
provide information on the largest spatial scales
\citep[see][]{Taylor+2003}.  In $Q$ and $U$, data from two
single-antenna surveys of the northern sky at $\sim$1.4~GHz, namely
the DRAO-26m survey \citep{Wolleben+2006} and the Effelsberg Medium
Latitude Survey, are added to the band-averaged ST data
\citep[see][]{Landecker+2008}.  All CGPS images presented in this
paper were produced using band-averaged data.

CGPS data calibration and processing procedures are described in
detail in \citet{Taylor+2003}.  Here we summarize for the reader the
general practice, emphasizing procedures related specifically to
polarization.  The complex antenna gains for the pointing centers in
each CGPS field are calibrated by observing a compact calibrator
source, either 3C~147 or 3C~295, at the start and end of each
observing session.  The polarization angle is calibrated using the
polarization calibrator source 3C~286.  Amplitude and phase variations
encountered during individual observing sessions (on time scales down
to $\sim$2~hr) are determined during the processing of the total
intensity data, and are applied also to the $Q$ and $U$ data.
Additional processing, needed to remove the effects of strong sources
both inside and outside the primary beam of the ST antennas, is
accomplished using routines developed especially for the DRAO ST
\citep[see][]{Willis1999}.  The instrumental polarization, which is
corrected on-axis in the sequence above, varies across the primary
beam of the ST antennas due to cross-polarization of the (nominally
orthogonal) receiver feeds and the effects of the feed support struts
\citep[see][]{Ng+2005}.  The result is ``leakage'' of unpolarized
radiation, seen in total intensity, into $Q$ and $U$.  We have
employed two different methods at DRAO to correct for the wide-field
instrumental polarization.  Both methods were used to calibrate the
various CGPS fields appearing to some degree in the
$2.5\arcdeg\times2.5\arcdeg$ region presented in this paper.  In the
first method, we derived the ``average'' leakage pattern in $Q$ and
$U$ across the primary beam of the ST antennas, and subtracted from
each of the $Q$ and $U$ images the ``leakage image'' for total
intensity into $Q$ and $U$, respectively \citep[see][]{Taylor+2003}.
In the second (and newer) method, we derived the leakage patterns for
each of the ST antennas separately, and subtracted the complex leakage
pattern created by each pair of antennas directly from the $Q$ and $U$
visibility data \citep[see][]{Reid+2008}.  The residual instrumental
polarization error after the on-axis and wide-field calibration is
similar for each method in each processed field, increasing from
$\sim$0.3\% root-mean-square (rms) at the field pointing center to
$\sim$1\% at the field edge ($\rho = 75\arcmin$).  The rms error is
reduced further in the mosaicing process.  The newer wide-field
correction significantly reduces artifacts in $Q$ and $U$ around
bright total-intensity sources (i.e., with flux densities
$\gtrsim$100~mJy).  No artifacts are seen above the estimated
$\sim$0.34~$\rm{mJy}\,\rm{beam}^{-1}$ ($\sim$0.086~K) noise level in
the $2.5\arcdeg\times2.5\arcdeg$ region of the CGPS presented in this
paper.

\section{RADIO POLARIZATION IMAGES OF PN \sh \label{results}}

\begin{figure}
%\centerline{\includegraphics[bb = 63 69 570 334,width=16cm,clip]{fig1.eps}}
\plotone{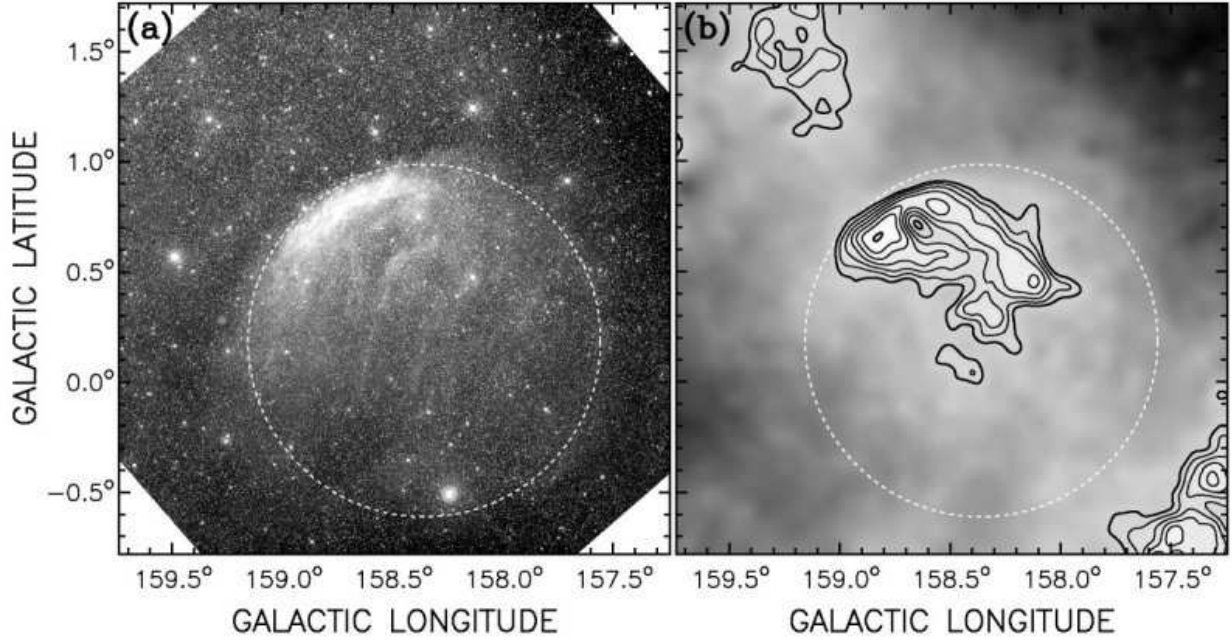}
\figcaption{Images of the $2.5\arcdeg\times2.5\arcdeg$ region centered
  approximately (see text) on the position of PN \sh\@ in $(a)$
  optical intensity at R-band (from the DSS) and $(b)$ total radio
  intensity at 1420~MHz.  Here and hereafter, images are presented in
  Galactic coordinates, with Galactic north up and Galactic east to
  the left.  The gray scale is in photon counts in $(a)$ and
  brightness temperature in $(b)$, with lighter shades indicating
  higher counts/temperatures.  The range of intensities in $(a)$ has
  been adjusted to highlight extended emission.  Point sources are
  removed in $(b)$ leaving extended emission with brightness
  temperatures in the range $4.57$--$4.92$\,K.  The contours drawn in
  $(b)$ accentuate higher brightness temperatures, and run from
  $4.86$\,K to $4.92$\,K in steps of $0.01$\,K.  The angular
  resolutions are $(a)$ $\sim$1\arcsec\@ and $(b)$ 5$\arcmin$
  (smoothed from $\sim$1\arcmin).  The dotted circle drawn on each
  image, and on each subsequent image, shows the approximate extent of
  the visible disk of \sh.
\label{radandoptimages}}
\end{figure}

In Figure~\ref{radandoptimages} we show images of the
$2.5\arcdeg\times2.5\arcdeg$ region around PN \sh\@ in both optical
intensity at R-band ($\lambda = 6570$\,nm) and total radio intensity
at 1420~MHz ($\lambda = 21$\,cm).  The optical image is taken from the
Digitized Sky Survey (DSS) and the radio image from the CGPS.  The
images are presented in Galactic coordinates and centered on the
position of the host white dwarf LS\,V\,$46\arcdeg21$ \citep[$l =
158.49\arcdeg$, $b = +0.47\arcdeg$; see][]{Kerber+2003}.  Note that
the center of the visible disk of \sh\@ is offset $\approx$24\arcmin\@
to the Galactic south-west of the white dwarf
\citep[see][]{TweedyMN1995}.  For the optical image, we adjusted the
range of intensities to highlight extended emission.  For the total
radio intensity image, we removed point sources leaving only extended
emission.  There is a clear enhancement in both images across much of
the face of \sh, relative to the surroundings, but the enhancement is
most intense along the (Galactic) north-eastern rim; i.e., in the
interaction region between \sh\@ and the ISM.  Treating the
enhancement in the radio image as thermal emission from the shell of
\sh, we can estimate the thermal electron density in the interaction
region of the PN (see \S~\ref{discuss}).

\begin{figure}
%\centerline{\includegraphics[bb = 63 69 570 334,width=16cm,clip]{fig2.eps}}
\plotone{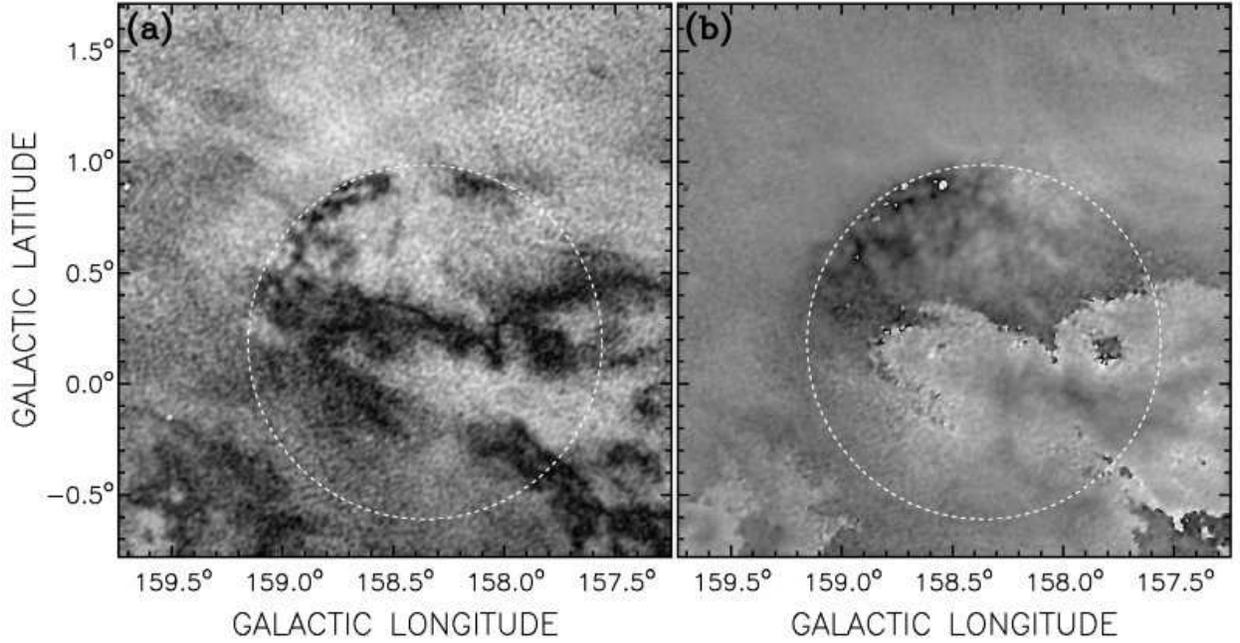}
\figcaption{Images in $(a)$ polarized intensity, $P =
  \sqrt{Q^2+U^2-(1.2\sigma)^2}$, and $(b)$ polarization angle,
  $\theta_P = \frac{1}{2}\arctan{U/Q}$.  The gray scale is in
  brightness temperature in $(a)$ and runs from $0$ to $0.67$\,K, with
  lighter shades indicating higher temperatures.  The gray scale in
  $(b)$ extends from $-90\arcdeg$ (black) to $+90\arcdeg$ (white).
  Note that abrupt black-to-white transitions in $(b)$ do not
  represent large changes in angle, since polarization angles of
  $-90\arcdeg$ and $+90\arcdeg$ are equivalent.  The resolving beam in
  each image is $1.31\arcmin \times 0.97\arcmin$ (full-width at
  half-maximum; FWHM) oriented at a position angle (east of north) of
  $-40\arcdeg$.
\label{piandpaimages}}
\end{figure}

In Figure~\ref{piandpaimages} we show the polarized intensity ($P =
\sqrt{Q^2+U^2-(1.2\sigma)^2}$, where the last term gives explicitly
the noise bias correction) and polarization angle ($\theta_P =
\frac{1}{2}\arctan{U/Q}$) images at 1420~MHz for the
$2.5\arcdeg\times2.5\arcdeg$ region in the CGPS around PN \sh.  The
polarization images contain several interesting features on a variety
of angular scales.  The most notable feature is a
low-polarized-intensity arc $\sim$0.15\arcdeg\@ wide and
$\sim$0.7\arcdeg\@ in length, coinciding with the north-east portion
of the visible disk of \sh.  The reduced intensity and distinct shape
of the arc indicate that its appearance is due to the effects of
Faraday rotation: specifically (1) localized beam depolarization
within the arc of the background diffuse synchrotron emission, and/or
(2) cancellation of the background emission, Faraday-rotated within
the arc, by foreground synchrotron emission.  For
background/foreground cancellation to play a significant role, the
polarized foreground emission must be a reasonable percentage of the
total polarized emission in the direction of \sh.  Galactic models for
synchrotron emission predict for the 129~pc foreground toward \sh\@
only $\sim$0.06~K at 1420~MHz \citep*[e.g.,][]{BeuermannKB1985}.  Even
if this emission is highly (i.e., $\sim$70\%) polarized, we expect
just $\sim$0.04~K of polarized emission in the foreground; i.e.,
$\lesssim$10\% of the $0.47 \pm 0.06$~K total polarized emission seen
outside the visible disk of \sh\@ (and at $b > +0.5\arcdeg$).  We
conclude, therefore, that beam depolarization plays a larger role in
reducing polarized emission over the arc than background/foreground
cancellation.  Structural variations observed within the arc in both
polarized intensity and polarization angle on angular scales down to
the resolution limit ($\sim$1\arcmin) further suggest that beam
depolarization is responsible for the appearance of this feature.
Since the length and location of the arc are very similar to the
optically bright rim denoting the PN-ISM interaction region, it would
seem there is a physical connection between the conditions and
processes in this region which give rise to enhanced optical emission
and those which lead to sharp gradients in $RM$.  We discuss the $RM$
structure within the arc in \S~\ref{arc}.

Aside from the prominent north-east arc, do we see other signatures of
\sh\@ in the polarization images?  The circle representing the visible
disk of \sh\@ in Figure~\ref{piandpaimages} draws our attention to two
suggestive details in the northern half ($b > +0.5\arcdeg$) of the
images: (1) the appearance of a second low-polarized-intensity ``arc''
$\sim$0.2\arcdeg\@ wide and $\sim$0.4\arcdeg\@ in length, located in
the north-west portion of the visible disk of \sh; and (2) the
increased range of polarization angles on the visible disk of \sh\@
compared to the surroundings.  The small-scale structural variations
in polarization angle within the north-west arc indicate that beam
depolarization is responsible to at least a moderate degree for the
reduced emission in this feature.  If this second arc is indeed
associated with \sh, then the conditions for sharp $RM$ gradients in
the shell of the PN may not be confined to the optically-identified
interaction region.  Moreover, a comparison of the range of
polarization angles seen inside the visible disk of \sh\@
($-82\arcdeg$ to $+27\arcdeg$, $\rm{rms}\approx17\arcdeg$) with those
seen outside ($-17\arcdeg$ to $+32\arcdeg$, $\rm{rms}\approx6\arcdeg$)
suggests that the conditions for moderate $RM$s are present throughout
the shell of \sh.  We present a simple model for the observed
polarization structures on the visible disk of \sh\@ in
\S~\ref{discuss}.

In contrast to the smaller-scale structures seen on the visible disk
of \sh\@ in the northern half of the images ($b > +0.5\arcdeg$), the
southern half of the images ($b < +0.5\arcdeg$) is dominated by
``bands'' of relatively low polarized intensity,
0.1\arcdeg--0.3\arcdeg\@ wide, which stretch approximately east-west
across the region.  The bands have no counterpart in total intensity.
The boundary between north and south is clearly marked in the
polarization angle image by a jagged line over which the angle changes
very rapidly.  On close inspection of the polarized intensity image,
this line corresponds to a narrow ($\sim$1\arcmin) channel within the
northernmost band of virtually zero polarized emission.  Changes in
the polarization angle across ``cells'' $<3\arcmin$ in size are seen,
to differing degrees, throughout the bands.  The bands appear to be
part of a large-scale complex which depolarizes the background diffuse
emission, most likely before it reaches the position of \sh.  In the
less likely scenario that the complex sits between the Sun and \sh,
any polarization signature imprinted on the background by \sh\@ is
lost.  In either case, the positioning of the bands on the sky south
of the north-east arc associated with \sh, and other apparent features
on the northernmost portion of the visible disk of the PN, would seem
to be fortuitous.

\subsection{Rotation-Measure Structure in the North-East Arc \label{arc}}

\begin{figure}
\centerline{\includegraphics[bb = 59 54 443 437,width=15cm,clip]{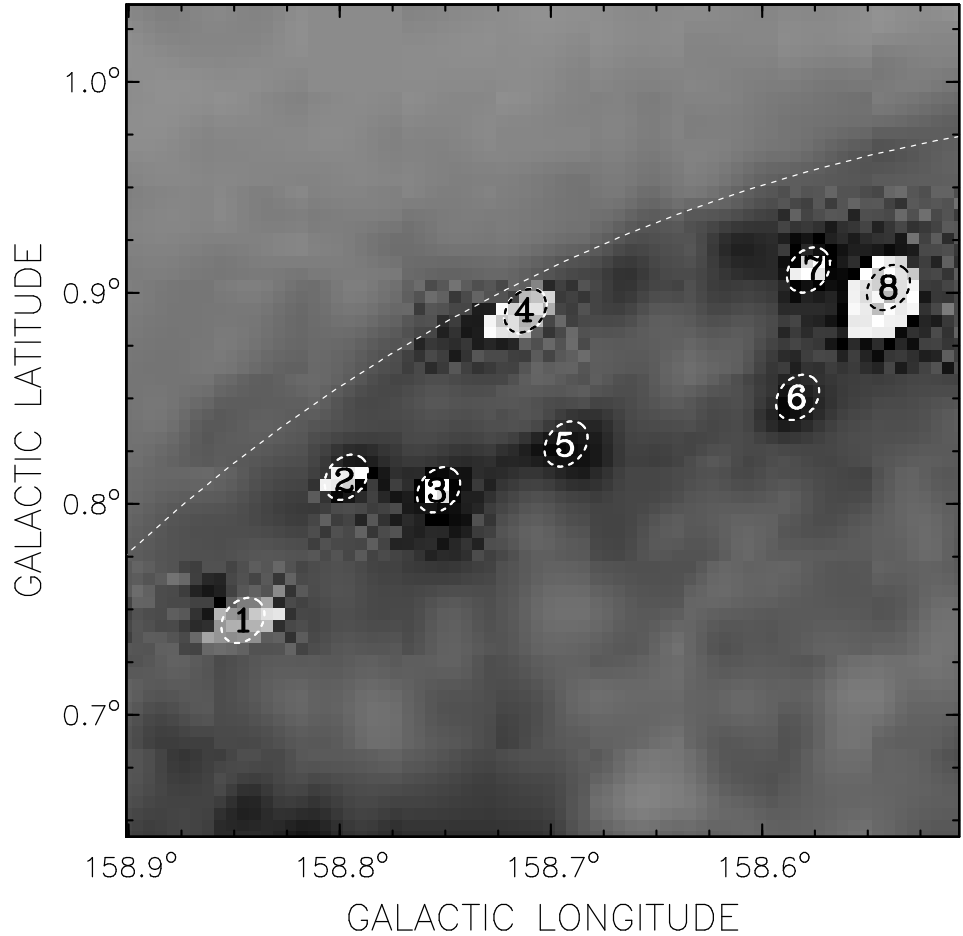}}
\figcaption{Polarization angle image zoomed to a
  $0.4\arcdeg\times0.4\arcdeg$ region around the north-east arc.  The
  gray scale is as described for Fig.\,\ref{piandpaimages}$b$.  The
  eight ``knots'' discussed in the text are labeled.  The ellipses
  (thick dashed lines) drawn around each knot represent the resolving
  beam (FWHM), and define the knot perimeters for which the
  polarization angles (see Table~\ref{knotangles}) are determined.
\label{zoompaimage}}
\end{figure}

The polarization angle of the relatively bright emission outside the
visible disk of \sh\@ (and at $b > +0.5\arcdeg$) has a mean value of
$+7$\arcdeg\@ and rms variations of only 6\arcdeg\@.  Along the
prominent north-east arc, the polarization angle is observed to change
rapidly across the perimeters of roughly elliptical ``knots.''  The
angle inside the perimeters changes more slowly and, indeed, plateaus
at the centers of the knots.  In Figure~\ref{zoompaimage} we show a
small $0.4\arcdeg\times0.4\arcdeg$ region in polarization angle around
the north-east arc and identify eight discrete knots.  We define as
the center of each knot the position of the pixel showing the maximum
clockwise (see below) deviation from the background ($+7\arcdeg \pm
6\arcdeg$) value.  In Table~\ref{knotangles} we give the mean value of
the polarization angle in each knot.  The mean was estimated over an
area corresponding to the area of the resolving beam (10 pixels, see
Figure~\ref{zoompaimage}), excluding, in the cases of knots 2 and 4,
pixels which differed from the 10-pixel mean by more than 2$\sigma$.
Table~\ref{knotangles} shows that the polarization angle of the
emission emerging from the knots is rotated significantly with respect
to the background emission.  The weighted mean polarization angle at
the centers of the knots is $+78\arcdeg \pm 22\arcdeg$.  If we assume
that emission with polarization angle $+7\arcdeg \pm 6\arcdeg$ is
incident on the far side of each knot, and for the moment ignore
foreground emission, then the incident emission is Faraday rotated in
the knots by $\Delta\theta = -109\arcdeg \pm 23\arcdeg$.  We infer
negative, i.e., clockwise, rotation by tracing polarization angles
from the outside edge of the arc to the center of any knot.  The trace
shows that the polarization angle (first) decreases through negatives
values.  At five of the eight knot perimeters, the polarization angle
jumps from $-90\arcdeg$ to $+90\arcdeg$, and then continues to
decrease to its center value.  Since foreground emission probably
cannot be ignored at the $\sim$10\% level, we must estimate the
maximum deviation expected in the observed polarization angle if, by
chance, the foreground emission is rotated 45\arcdeg\@ relative to the
emission emerging from the knots.  (Note that foreground emission
rotated 90\arcdeg\@ relative to the background leads to a maximum
reduction in polarized intensity, but no net rotation in polarization
angle.)  Assuming complete beam depolarization at the perimeters of at
least some of the knots, in particular knots 1 and 4, we estimate the
polarized foreground emission to be $0.046 \pm 0.012$~K, consistent
with the values predicted by Galactic synchrotron models.  Using
$0.177 \pm 0.044$~K for the mean observed (i.e., emerging plus
foreground) polarized emission at the centers of the knots (see
Table~\ref{knotangles}), we estimate a maximum deviation of 7\arcdeg.
Adding this in quadrature to the 23\arcdeg\@ statistical uncertainty
gives a standard error for the measured rotation through the knots of
24\arcdeg.  For a center wavelength of 21.12~cm (see \S~\ref{obs}), a
rotation of $-109\arcdeg \pm 24\arcdeg$ gives (via Equation 1) $RM =
-43 \pm 10$~$\rm{rad}\,\rm{m}^{-2}$.

The emission emerging from the centers of the knots, where the
polarization angles plateau, is likely higher than the $0.177 \pm
0.044$~K value given above, but polarization-angle variations over the
resolving beam lead to reduced polarized intensity even inside the
knot perimeters.  While these variations are reflected in the range of
angles observed in each knot (see Table~\ref{knotangles}), we
nevertheless believe our estimated mean $RM$ reflects a real
systematic rotation of the background emission as it passes through
the knots.  The consistent clockwise rotation of the polarization
angle observed in moving from outside the north-east arc toward the
center of any knot strengthens this assertion.

We tried to estimate the $RM$ through the knots using the four-band
data from the ST but failed.  Given the typical uncertainty in
polarization angle in the band-averaged image, and noting that $RM
\approx -43$~$\rm{rad}\,\rm{m}^{-2}$ gives a difference in rotation
angle of only $\sim$5\arcdeg\@ over 27.5~MHz ($\Delta\lambda =
0.41$~cm), the failure is not surprising.

\section{DISCUSSION \label{discuss}}

We can derive the line-of-sight component of the magnetic field
through the knots in the north-east arc of \sh\@ using $-43 \pm 10\ 
\rm{rad}\,\rm{m}^{-2}$ as an estimate of the $RM$ in the knots, and
using estimates of the thermal electron density in and path length
through the interaction region (see Equation 2).  The electron density
in \sh\@ can be calculated from emission measure ($EM =
\int{{n_e}^2\,dl}$) determinations, made independently at optical and
radio wavelengths.  Based on their measured H$\alpha$ intensity and
gas temperature ($T_e = 9400 \pm 1100$~K), \citet{Reynolds1985}
estimates a mean value for the emission measure over \sh\@ of $EM
\approx 42$~$\rm{cm}^{-6}$\,pc.  Using a brightness temperature of
$T_b = 0.11 \pm 0.02$~K for the thermal radio emission in the
interaction region of \sh\@ (obtained via comparison of on-source and
off-source temperatures in Figure~\ref{radandoptimages}$b$), and the
same gas temperature, we estimate a value for the interaction region
of $EM = 69 \pm 13$~$\rm{cm}^{-6}$\,pc.  If we assume for the moment
that electrons are uniformly distributed over the approximately
spherical volume of \sh, and use an average path length through the
sphere of $\Delta l = \frac{4}{3} R_{\rm{PN}} \approx 2.5$~pc, then
the optically-determined $EM$ gives a mean electron density over \sh\@
of $n_e \approx 4.1$~$\rm{cm}^{-3}$.  In some contrast, the
radio-intensity-determined $EM$ gives, for a path length\footnote{The
  path length $\Delta l = 1.1 \pm 0.3$~pc corresponds to the mean of
  the line-of-sight chord lengths through a 1.9-pc radius sphere at
  the positions of the eight knots.} through the interaction region of
$\Delta l = 1.1 \pm 0.3$~pc, $n_e = 7.9 \pm 1.3$~$\rm{cm}^{-3}$.  The
factor $\sim$2 increase in the electron density in the interaction
region compared to the mean value over the entire PN is reasonable,
since material is stacking up in the interaction region
\citep[see][]{TweedyMN1995}.  However, $n_e = 7.9 \pm
1.3$~$\rm{cm}^{-3}$ still represents a mean over the interaction
region.  The H$\alpha$ images of \citet{TweedyMN1995} show small-scale
filamentary structures in the interaction region with localized factor
1.5--2 enhancements in $EM$ relative to the mean.  The filaments
correspond approximately in both location and size to the
polarization-angle knots.  Since we cannot confirm the physical
association between the filaments and the knots, we conservatively
assume an $EM$-enhancement of $1.5 \pm 0.5$, and estimate the electron
density in the knots to be $n_e = 9.7 \pm 2.3$~$\rm{cm}^{-3}$.  Using
$RM = -43 \pm 10$~$\rm{rad}\,\rm{m}^{-2}$, $n_e = 9.7 \pm
2.3$~$\rm{cm}^{-3}$ and $\Delta l = 1.1 \pm 0.3$~pc, we derive a
line-of-sight magnetic field through the knots in the interaction
region of $B_{\|} = 5.0 \pm 2.0$~$\mu\rm{G}$.  Since the $RM$ is
negative, this field is directed into the plane of the sky.

\subsection{An ISM Origin for the Magnetic Field in the Shell of \sh \label{ismfield}}

Is a $\sim$5~$\mu\rm{G}$ line-of-sight magnetic field reasonable for
the ISM around \sh?  Since there is no direct measurement of the ISM
magnetic field around \sh, we estimate the local field from what is
known generally about the Galactic magnetic field.  The Galactic
magnetic field is concentrated in the disk and has two components
\citep[see, e.g.,][]{Beck+1996}: a large-scale or regular component
($B_{reg}$), which follows the spiral arms, and a small-scale or
random component ($B_{ran}$).  $B_{reg}$ in the local spiral arm is
found, using polarized radio sources and the polarization of
starlight, to be directed toward $l \approx 85\arcdeg$
\citep[e.g.,][]{RandL1994,Heiles1996a,BrownT2001}; i.e., clockwise as
viewed from the Galactic north pole.  Fluctuation cell sizes for
$B_{ran}$ are estimated to be 50--100~pc
\citep[see][]{RandK1989,OhnoS1993}.  The ratio of the strengths of the
random and regular components of the Galactic field,
$B_{ran}/B_{reg}$, can be obtained directly from starlight
polarization data and synchrotron polarization data using the model
presented in \citet{Burn1966}.  The starlight polarization data of
\citet{Fosalba+2002} give for a large sample of stars covering all
Galactic longitudes $B_{ran}/B_{reg} \approx 1.3$.  Using stars from
the sample of \citet{MathewsonF1970} in the range $120\arcdeg < l <
180\arcdeg$, and a modified version of the Burn model,
\citet{Heiles1996b} finds $B_{ran}/B_{reg} \approx 1.5$.  For
synchrotron emission just north of the visible disk of \sh\@ ($l =
158.5\arcdeg$), we measure a fractional linear polarization of $p =
0.27 \pm 0.03$, close to the maximum value found by
\citet{Spoelstra1984} for the diffuse emission in the Galactic plane.
With a spectral index $\alpha = -0.44 \pm 0.04$ ($S \propto
\nu^{\alpha}$) between 408~MHz and 1420~MHz for the synchrotron
emission in the CGPS region around \sh, we get for the intrinsic value
of the fractional linear polarization $p_{max} = 0.68 \pm 0.01$
\citep[see][]{GinzburgS1965}, and thus obtain $B_{ran}/B_{reg} = 1.51
\pm 0.13$, consistent with the \citet{Heiles1996b} starlight estimate.
Using a value $B_{tot} = 4.2$~$\mu\rm{G}$ for the average azimuthal
field strength ($B_{tot}^2 = B_{reg}^2 + B_{ran}^2$) in the local arm
\citep{Heiles1996b} and $B_{ran}/B_{reg} = 1.51$, we estimate $B_{reg}
\approx 2.3$~$\mu\rm{G}$ and $B_{ran} \approx 3.5$~$\mu\rm{G}$.  The
maximum magnetic field at any point in the local arm is then achieved
if, by chance alignment, the random field lies parallel to the regular
field; i.e., $B_{max} = B_{reg} + B_{ran} \approx 5.8$~$\mu\rm{G}$.
At the longitude of \sh, both the average field ($B_{tot} =
4.2$~$\mu\rm{G}$) and maximum possible field ($B_{max} =
5.8$~$\mu\rm{G}$) lie largely in the plane of the sky, and run from
Galactic east to west.  The maximum field along the line-of-sight,
where $B_{\|reg} \approx 0.6$~$\mu\rm{G}$, is $B_{\|max} = B_{\|reg} +
B_{ran} \approx 4.1$~$\mu\rm{G}$, directed into the plane of the sky.

In light of this brief overview, we conclude that an intrinsic
$\sim$5~$\mu\rm{G}$ line-of-sight magnetic field in the ISM at the
position of \sh\@ is unlikely.  Nevertheless, our observations can be
used to comment further on the structure of a proposed ISM field in
the interaction region as well as other locations in the shell of \sh.
Our estimate of the line-of-sight magnetic field in the interaction
region is based on the maximum $RM$ as seen through knots in our
polarization angle image.  The $RM$s outside the knots are apparently
much lower.  The sharp $RM$ gradients over the knot perimeters must be
the result of either a rapid change in the electron density or the
line-of-sight magnetic field, or both.  As we previously noted, the
polarization-angle knots appear to be associated with narrow H$\alpha$
filaments observed in the interaction region.  The sharp edges of the
filaments, which denote a rapid change in $EM$ (and thus electron
density), naturally explain $RM$ gradients across knot perimeters.
The magnetic field need not change significantly across the
interaction region.  Realistically, however, the magnetic field is
probably affected by turbulence in the hot gas (see
\S~\ref{simplemodel}).

If we look west of the north-east arc, toward the center of the
visible disk of \sh, we continue to see polarization angles
significantly different from the $+7\arcdeg \pm 6\arcdeg$ observed
outside the PN (see Figure~\ref{piandpaimages}$b$).  Though we don't
see prominent structures in this ``interior'' region, we do see some
localized polarization-angle structures.  These structures are roughly
coincident with low-level enhancements in H$\alpha$ and \ion{N}{2}
\citep[see][]{TweedyMN1995} and total radio intensity (see
Figure~\ref{radandoptimages}$b$).  Localized electron-density
enhancements may therefore be responsible for both the knots in the
interaction region and the more extended structures seen across the
western portion of the face of \sh.  Indeed, these two apparently
different structures may arise from similar underlying structures,
seen edge-on in the case of the knots, and face-on in the case of the
extended structures \citep[see, e.g.,][]{TweedyMN1995}.  The increased
path length through the shell in the interaction region would explain,
at least in part, why the $EM$s and $RM$s in the filaments/knots are
larger than those in the extended structures across the face.  A
decrease in the line-of-sight component of the magnetic field, moving
west from the north-east edge of \sh\@ toward the center of the
visible disk, could also account for some of the difference (see
\S~\ref{simplemodel}).

A second low-polarized-intensity arc appears at the north-west edge of
the visible disk of \sh\@ (see Figure~\ref{piandpaimages}$a$).
Small-scale variations in polarization angle within this arc indicate,
as in the north-east arc, the presence of sharp $RM$ gradients.
However, unlike the north-east arc, there are no plateaus (i.e.,
multi-pixel regions of roughly constant polarization angle) over which
we can confidently estimate some deviation from the outside $+7\arcdeg
\pm 6\arcdeg$.  Consequently, we have no means of estimating the
magnitude of the $RM$ through this arc.  Nevertheless, there is some
indication of the sign of the $RM$.  Moving south from the edge of the
north-west arc, the polarization angle (on average) increases,
implying positive $RM$s.  To substantiate this finding, we broke the
north-west arc into three north-south slices, and used the approach of
\citet{WollebenR2004} to estimate $RM$ together with three other
parameters (degree of depolarization, foreground polarized intensity
and foreground polarization angle).  We found positive $RM$s for each
slice, even when we varied the other parameters away from their
``best-fit'' values.  The positive $RM$s indicate that the magnetic
field in the north-west arc is directed out of the plane of the sky.
If the north-west arc is associated with \sh, and the ISM field is
responsible for the observed $RM$s in both the north-east and
north-west arcs, then the intrinsic field must be deflected around the
PN.

\subsubsection{A simple model for the ISM magnetic field around \sh \label{simplemodel}}

\begin{figure}
\centerline{\includegraphics[bb = 0 1 537 661,width=13cm,clip]{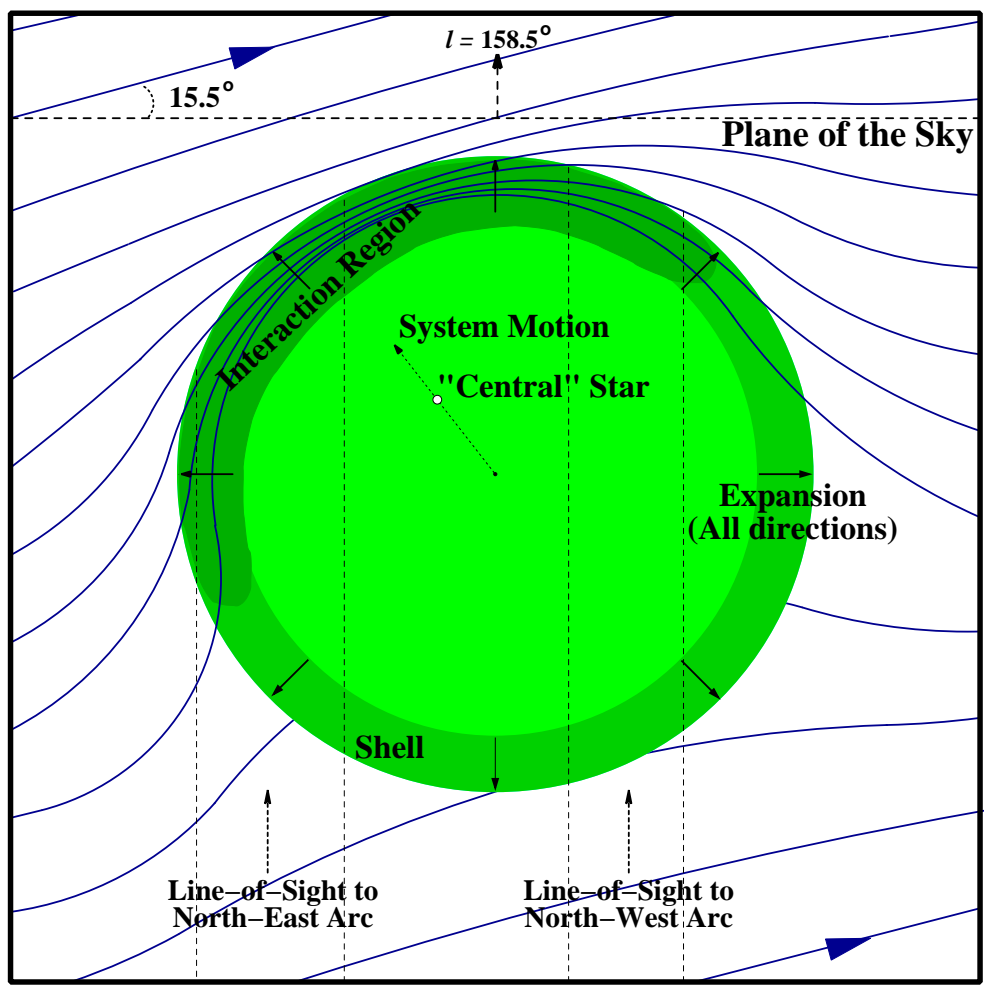}}
\figcaption{Simple model showing the interaction between \sh\@ and the
  ISM magnetic field.  The perspective is that of an observer sitting
  above the Galactic plane and looking down on the center of the PN,
  with $l = 158.5\arcdeg$ directed upward.  The PN is expanding in all
  directions.  The motion of the PN system projected onto the Galactic
  plane is indicated.  The host white dwarf (``central'' star) is
  offset from the center of the PN toward the interaction region,
  where the ISM magnetic field (solid lines) is compressed and
  deflected around the shell of the PN.  Magnetic field lines which
  appear to stop at the edge of the PN actually slide on the surface
  either above or below the slice shown.  The intrinsic ISM field is
  inclined 15.5\arcdeg\@ to the plane of the sky.  The lines-of-sight
  for an Earthbound observer to the north-east and north-west arcs are
  indicated.
\label{modelpic}}
\end{figure}

For the ISM magnetic field to simultaneously account for the negative
$RM$s observed in the north-east arc and the positive $RM$s observed
in the north-west arc, the intrinsic field must bend significantly
around the shell of \sh\@ such that it has a $\sim$5~$\mu\rm{G}$
line-of-sight component into the sky on the east edge of the PN and a
non-zero line-of-sight component out of the sky on the west edge.
This is exactly what we might expect in the following scenario (see
Figure~\ref{modelpic}): The intrinsic ISM field around \sh\@ is
described by the 4.2~$\mu\rm{G}$ azimuthal component of the Galactic
magnetic field \citep[see][]{Heiles1996b}, which, at $l =
158.5\arcdeg$, runs in the local arm from Galactic east to west and
intersects the plane of the sky at $15.5\arcdeg \pm 4\arcdeg$
\citep{BrownT2001}.  The intrinsic field is compressed and deflected
by the expanding and moving PN, since it can diffuse only slowly into
the partially ionized shell \citep[see][]{SokerD1997}.  The
three-dimensional motion of \sh\@ is fully characterized by the
(Galactic) north-west-directed motion of the host white dwarf
\citep{CudworthR1985,TweedyMN1995} and a line-of-sight motion into the
plane of the sky (see below).  The motion in the plane of the sky is
of less importance for our observations than the motion along the
line-of-sight, though the full three-dimensional picture is important
for interpreting the alignment of the filaments in the interaction
region and wider structures across the face of \sh\@
\citep[see][]{TweedyMN1995}.  At the far side of \sh, the
line-of-sight motion drags the intrinsic field away from the observer.
The result is a deflected field around \sh\@ which has on the east
edge of the PN a line-of-sight component directed into the sky and on
the west edge a line-of-sight component directed out of the sky.  In
the center portion of the PN, the field lies largely in the plane of
the sky.  The line-of-sight component of the field on the east edge is
slightly larger in strength than the intrinsic field itself, while
that on the west side is lower.  We estimate the line-of-sight motion
of \sh\@ by consulting the Wisconsin H$\alpha$ Mapper (WHAM) survey
\citep{Haffner+2003}.  The WHAM data show that the H$\alpha$ emission
from \sh\@ peaks at velocity $+5 \pm 1$~$\rm{km}\,\rm{s}^{-1}$
relative to the local standard of rest.  (Note that the sign for
velocity is opposite that of $RM$; i.e., a positive velocity signifies
motion into the plane of sky while a positive $RM$ signifies a
magnetic field out of the plane of the sky).  The emission surrounding
\sh\@ peaks at velocities near zero, indicating that the $+5 \pm
1$~$\rm{km}\,\rm{s}^{-1}$ is indeed relative to the surrounding ISM.

The magnetic field described by our model represents only the
``smooth'' component of the deflected ISM field.  The magnetic field
in the shell of the PN will also have a turbulent component due to
motions in the hot gas.  The turbulent component contributes in part
to the beam depolarization we observe in the interaction region.  The
smooth component is responsible for the systematic $RM$ observed
through the knots.

\subsubsection{Are old PNe good potential probes of the magnetized ISM? \label{potentialasprobe}}

We have asked in this section whether or not the magnetic field
derived from the observed $RM$s is reasonable for the ISM around \sh.
If we were, instead, to concede that the derived field is native to
the ISM, then we could ask the question: Are old PNe, such as \sh,
good probes of the intrinsic ISM field?  The simple qualitative model
presented above for \sh\ suggests that we can learn something about
the intrinsic field surrounding this PN.  With a more comprehensive
(three-dimensional) model of the PN-ISM interaction, it may be
possible to work out quite accurately the strength and orientation of
the intrinsic field.  Since a large percentage of old PNe show PN-ISM
interactions similar to \sh\@ \citep[see][]{TweedyK1996}, it is
perhaps reasonable to assume that the conditions for detectable
Faraday rotation are present in the shells of many old PNe in the
nearby Galaxy.  Given a good model of the interaction in each case, it
should then be possible to determine the intrinsic field at many
locations.

There are two significant drawbacks to consider before we declare old
PNe good potential probes of the magnetized ISM: (1) It may not be
possible in many cases to construct a good model of the PN-ISM
interaction, due either to large uncertainties in the physical
parameters (e.g., ISM and PN particle densities, space velocity of the
PN) or to the overall complexity of the interaction.  The qualitative
model presented above for \sh\@ does not comment on either the degree
to which the intrinsic field around the PN is compressed or the
maximum angle with which the intrinsic field is deflected.  Both of
these quantities are necessary in order to more accurately determine
the intrinsic field from the line-of-sight field.  Three-dimensional
magneto-hydrodynamic (MHD) simulations can perhaps be used to
demonstrate the sensitivity of the intrinsic-field determination to
various parameters. (2) The Faraday signature of the PN is at the
mercy of fluctuations in the warm ionized ISM as well as turbulent
structures (e.g., \ion{H}{2} regions) which may lie along the
line-of-sight.  The dark bands that run through the approximate
midpoint of \sh\@ (almost) completely depolarize the diffuse
background emission.  If \sh\@ were located $\sim$0.5\arcdeg\@ south
of its actual position, then the distinct signature of the north-east
arc would be destroyed.  Given the prevalence of turbulence in the
ISM, this point is a significant concern.  In fact, of the six PNe in
the CGPS region known to interact with the ISM, only two, including
\sh, are seen in polarization.  (The other, namely DeHt\,5, is the
subject of a subsequent paper.)  Targeted polarimetric observations of
old PNe at sub-arcminute resolution and at multiple frequencies in the
range 1--3~GHz are necessary to better establish the potential of
these objects as good probes of the magnetized ISM.

\subsection{A Stellar Origin for the Magnetic Field in the Shell of \sh \label{stellarfield}}

Can a $\sim$5~$\mu\rm{G}$ magnetic field in the interaction region be
attributed to the host white dwarf near the center of \sh?  The
magnetic fields of white dwarfs have been measured via
spectropolarimetric observations of optical absorption lines, but only
recently have the observations had the sensitivity to detect kilogauss
fields.  The studies thus far have focused on either fully evolved
(compact) white dwarfs \citep{AznarCuadrado+2004} or central stars of
relatively young PNe which are still transitioning to white dwarfs
\citep*{JordanWO2005}.  In the case of evolved white dwarfs,
\citet{AznarCuadrado+2004} found for a sample of 12 stars only three
which had detectable magnetic fields in the range 2--4~kG, a detection
rate of 25\%.  On the other hand, \citet{JordanWO2005} found for each
of a selection of four transition stars magnetic fields of 1--3~kG, a
detection rate of 100\%.  Though the number statistics for both cases
are relatively poor, this pair of results suggests that the magnetic
fields of transition stars are present not in their degenerate cores
but rather their extended envelopes, since magnetic flux is apparently
lost during white dwarf evolution (i.e., during collapse from stellar
radii in the \citealt{JordanWO2005} sample of 0.14--0.3\,$R_{\sun}$ to
white dwarf radii of $\approx$0.012\,$R_{\sun}$).  The magnetic fields
of the central stars of old PNe have not been measured.  Given their
intermediate radius, e.g., 0.05\,$R_{\sun}$ for the ``central'' star
in \sh\@ \citep[based on the luminosity and effective temperature
given in][]{Rauch+2007}, measurements of the magnetic fields of the
central stars of old PNe could lead to an improved understanding of
field evolution in white dwarfs.  We point out for completeness that a
small fraction ($\sim$10\%) of white dwarfs are observed to have
magnetic fields at the 1~MG level or higher
\citep[see][]{Liebert+2003}, but these stars tend to have masses
($\approx$0.9\,$M_{\sun}$) much higher than typical white dwarf masses
(0.48--0.65\,$M_{\sun}$), and may come from magnetized progenitors
such as peculiar (Ap) stars \citep[see][]{Liebert1998,Liebert+2003}.

For present purposes, we assume the magnetic field in the envelope
around the contracting ``central'' star in \sh\@ to be accurately
represented by the magnetic field ($B_{\rm{avg}} = 1.8$~kG) measured
at the radii ($r_{\rm{avg}} = 0.21\,R_{\sun}$) of the central stars in
the \citet{JordanWO2005} sample.  With some knowledge of the
large-scale magnetic field geometry, we can then estimate the field at
large radii; namely, in the shell of \sh.  Unfortunately, at this
time, neither observations nor theory form a complete picture of the
magnetic fields in PNe.  Magnetic field measurements of maser spots in
precursor (AGB) circumstellar envelopes suggest a radial dependence of
the field \citep[$B \sim r^{-2}$; see][]{VlemmingsDL2002}, while
measurements for the supergiant VX\,Sgr show a poloidal dependence
\citep*[$B \sim r^{-3}$; see][]{VlemmingsLD2005}.  In contrast, the
geometry of filamentary structures observed by \citet{HugginsM2005} in
three PNe, as well as measurements by \citet*{VlemmingsDI2006} of the
magnetic field structure in the jet emanating from AGB star W43A,
suggest the dominance of toroidal fields ($B \sim r^{-1}$), consistent
with the theoretical framework of \citet{ChevalierL1994}.  If either
radial or poloidal geometries hold for \sh, then the magnetic field in
the interaction region ($\approx$1.0~pc from the host white dwarf)
would fall well below our $RM$-estimated $\sim$5~$\mu\rm{G}$
line-of-sight field.  On the other hand, if a toroidal field holds,
then the magnetic field in the interaction region could be
$\sim$8~$\mu\rm{G}$.  Given the spherical symmetry of the shell of
\sh, a large-scale toroidal magnetic field for this PN, invoked
generally to explain non-spherical (e.g., bipolar, elliptical)
symmetries in young PNe, is unlikely.  However, localized enhancements
of the internal magnetic field, due to compression in dense knots or
filaments \citep[see, e.g.,][]{Soker2002,SokerK2003}, are possible.
Thus we cannot completely dismiss the possibility of a
$\sim$5~$\mu\rm{G}$ internal field in the shell of \sh.  Detailed MHD
simulations need to be done in order to better understand the magnetic
field geometry in PNe.

\section{CONCLUSIONS \label{concl}}

Here we give a summary of our results and conclusions:

1. We presented 1420~MHz polarization images for the
$2.5\arcdeg\times2.5\arcdeg$ region in the CGPS around the PN \sh.

2. A low-polarized-intensity arc, $0.2\arcdeg \times 0.7\arcdeg$ in
size, appears in the north-east portion of the visible disk of \sh.
The arc is coincident with the optically-identified interaction region
between the PN and the ISM.

3. A second low-polarized-intensity arc appears in the north-west
portion of the visible disk of \sh.

4. The north-east arc contains structural variations down to the
$\sim$1\arcmin\@ resolution limit in both polarized intensity and
polarization angle.  Several polarization-angle ``knots'' appear along
the arc.

5. Via comparison of the polarization angles at the centers of the
knots in the north-east arc and the mean polarization angle outside
\sh\@ (and above $b \simeq +0.5\arcdeg$), we estimated the $RM$
through the knots to be $-43 \pm 10\ \rm{rad}\,\rm{m}^{-2}$.

5. Using this estimate for the $RM$ and an estimate of the electron
density in the shell of \sh, we derived a line-of-sight magnetic field
in the interaction region of $5.0 \pm 2.0$~$\mu\rm{G}$.

6. We believe it more likely the derived magnetic field is
interstellar than stellar, though we cannot completely dismiss the
latter possibility.  We interpret our observations via a simple model
which qualitatively describes the ISM magnetic field around \sh.

7. It is unclear whether old PNe like \sh\@ could be useful probes of
the magnetized ISM.  Targeted polarimetric observations at high
resolution ($<$1\arcmin), and possibly at multiple frequencies in the
range 1--3~GHz, may help separate the signatures of more PNe from the
turbulent ISM.

\acknowledgements

ACKNOWLEDGMENTS.  We thank an anonymous referee for a constructive
review of the paper and for comments helpful in the preparation of the
final manuscript.  R.R.R.\@ would like to thank Maik Wolleben for
applying his Faraday screen model to our data and for insightful
discussions.  The Canadian Galactic Plane Survey is a Canadian project
with international partners, and is supported by a grant from NSERC.
The Dominion Radio Astrophysical Observatory is operated as a national
facility by the National Research Council of Canada.  This research is
based in part on observations with the 100-m telescope of the MPIfR at
Effelsberg.  The Second Palomar Observatory Sky Survey (POSS-II) was
made by the California Institute of Technology with funds from the
National Science Foundation, the National Geographic Society, the
Sloan Foundation, the Samuel Oschin Foundation, and the Eastman Kodak
Corporation.  The Wisconsin H-Alpha Mapper is funded by the National
Science Foundation.

% ** Bibliography here **

\bibliographystyle{apj}
\bibliography{ms}

\begin{thebibliography}{59}
\expandafter\ifx\csname natexlab\endcsname\relax\def\natexlab#1{#1}\fi

\bibitem[{{Aznar Cuadrado} {et~al.}(2004){Aznar Cuadrado}, {Jordan},
  {Napiwotzki}, {Schmid}, {Solanki}, \& {Mathys}}]{AznarCuadrado+2004}
{Aznar Cuadrado}, R., {Jordan}, S., {Napiwotzki}, R., {Schmid}, H.~M.,
  {Solanki}, S.~K., \& {Mathys}, G. 2004, \aap, 423, 1081

\bibitem[{{Bains} {et~al.}(2003){Bains}, {Bryce}, {Mellema}, {Redman}, \&
  {Thomasson}}]{Bains+2003}
{Bains}, I., {Bryce}, M., {Mellema}, G., {Redman}, M.~P., \& {Thomasson}, P.
  2003, \mnras, 340, 381

\bibitem[{{Beck} {et~al.}(1996){Beck}, {Brandenburg}, {Moss}, {Shukurov}, \&
  {Sokoloff}}]{Beck+1996}
{Beck}, R., {Brandenburg}, A., {Moss}, D., {Shukurov}, A., \& {Sokoloff}, D.
  1996, \araa, 34, 155

\bibitem[{{Beuermann} {et~al.}(1985){Beuermann}, {Kanbach}, \&
  {Berkhuijsen}}]{BeuermannKB1985}
{Beuermann}, K., {Kanbach}, G., \& {Berkhuijsen}, E.~M. 1985, \aap, 153, 17

\bibitem[{{Brown} \& {Taylor}(2001)}]{BrownT2001}
{Brown}, J.~C. \& {Taylor}, A.~R. 2001, \apjl, 563, L31

\bibitem[{{Burn}(1966)}]{Burn1966}
{Burn}, B.~J. 1966, \mnras, 133, 67

\bibitem[{{Chevalier} \& {Luo}(1994)}]{ChevalierL1994}
{Chevalier}, R.~A. \& {Luo}, D. 1994, \apj, 421, 225

\bibitem[{{Cudworth} \& {Reynolds}(1985)}]{CudworthR1985}
{Cudworth}, K. \& {Reynolds}, R.~J. 1985, \pasp, 97, 175

\bibitem[{{Fosalba} {et~al.}(2002){Fosalba}, {Lazarian}, {Prunet}, \&
  {Tauber}}]{Fosalba+2002}
{Fosalba}, P., {Lazarian}, A., {Prunet}, S., \& {Tauber}, J.~A. 2002, \apj,
  564, 762

\bibitem[{{Gaensler} {et~al.}(2001){Gaensler}, {Dickey}, {McClure-Griffiths},
  {Green}, {Wieringa}, \& {Haynes}}]{Gaensler+2001}
{Gaensler}, B.~M., {Dickey}, J.~M., {McClure-Griffiths}, N.~M., {Green}, A.~J.,
  {Wieringa}, M.~H., \& {Haynes}, R.~F. 2001, \apj, 549, 959

\bibitem[{{Ginzburg} \& {Syrovatskii}(1965)}]{GinzburgS1965}
{Ginzburg}, V.~L. \& {Syrovatskii}, S.~I. 1965, \araa, 3, 297

\bibitem[{{Gray} {et~al.}(1999){Gray}, {Landecker}, {Dewdney}, {Taylor},
  {Willis}, \& {Normandeau}}]{Gray+1999}
{Gray}, A.~D., {Landecker}, T.~L., {Dewdney}, P.~E., {Taylor}, A.~R., {Willis},
  A.~G., \& {Normandeau}, M. 1999, \apj, 514, 221

\bibitem[{{Haffner} {et~al.}(2003){Haffner}, {Reynolds}, {Tufte}, {Madsen},
  {Jaehnig}, \& {Percival}}]{Haffner+2003}
{Haffner}, L.~M., {Reynolds}, R.~J., {Tufte}, S.~L., {Madsen}, G.~J.,
  {Jaehnig}, K.~P., \& {Percival}, J.~W. 2003, \apjs, 149, 405

\bibitem[{{Harris} {et~al.}(2007){Harris}, {Dahn}, {Canzian}, {Guetter},
  {Leggett}, {Levine}, {Luginbuhl}, {Monet}, {Monet}, {Pier}, {Stone},
  {Tilleman}, {Vrba}, \& {Walker}}]{Harris+2007}
{Harris}, H.~C., {Dahn}, C.~C., {Canzian}, B., {Guetter}, H.~H., {Leggett},
  S.~K., {Levine}, S.~E., {Luginbuhl}, C.~B., {Monet}, A.~K.~B., {Monet},
  D.~G., {Pier}, J.~R., {Stone}, R.~C., {Tilleman}, T., {Vrba}, F.~J., \&
  {Walker}, R.~L. 2007, \aj, 133, 631

\bibitem[{{Haverkorn} {et~al.}(2006{\natexlab{a}}){Haverkorn}, {Gaensler},
  {Brown}, {Bizunok}, {McClure-Griffiths}, {Dickey}, \&
  {Green}}]{Haverkorn+2006a}
{Haverkorn}, M., {Gaensler}, B.~M., {Brown}, J.~C., {Bizunok}, N.~S.,
  {McClure-Griffiths}, N.~M., {Dickey}, J.~M., \& {Green}, A.~J.
  2006{\natexlab{a}}, \apjl, 637, L33

\bibitem[{{Haverkorn} {et~al.}(2006{\natexlab{b}}){Haverkorn}, {Gaensler},
  {McClure-Griffiths}, {Dickey}, \& {Green}}]{Haverkorn+2006b}
{Haverkorn}, M., {Gaensler}, B.~M., {McClure-Griffiths}, N.~M., {Dickey},
  J.~M., \& {Green}, A.~J. 2006{\natexlab{b}}, \apjs, 167, 230

\bibitem[{{Haverkorn} {et~al.}(2003{\natexlab{a}}){Haverkorn}, {Katgert}, \&
  {de Bruyn}}]{HaverkornKd2003a}
{Haverkorn}, M., {Katgert}, P., \& {de Bruyn}, A.~G. 2003{\natexlab{a}}, \aap,
  403, 1031

\bibitem[{{Haverkorn} {et~al.}(2003{\natexlab{b}}){Haverkorn}, {Katgert}, \&
  {de Bruyn}}]{HaverkornKd2003c}
---. 2003{\natexlab{b}}, \aap, 404, 233

\bibitem[{{Haverkorn} {et~al.}(2004{\natexlab{a}}){Haverkorn}, {Katgert}, \&
  {de Bruyn}}]{HaverkornKd2004b}
---. 2004{\natexlab{a}}, \aap, 427, 169

\bibitem[{{Haverkorn} {et~al.}(2004{\natexlab{b}}){Haverkorn}, {Katgert}, \&
  {de Bruyn}}]{HaverkornKd2004a}
---. 2004{\natexlab{b}}, \aap, 427, 549

\bibitem[{{Heiles}(1996{\natexlab{a}})}]{Heiles1996b}
{Heiles}, C. 1996{\natexlab{a}}, in Astronomical Society of the Pacific
  Conference Series, Vol.~97, Polarimetry of the Interstellar Medium, ed. W.~G.
  {Roberge} \& D.~C.~B. {Whittet}, 457--+

\bibitem[{{Heiles}(1996{\natexlab{b}})}]{Heiles1996a}
{Heiles}, C. 1996{\natexlab{b}}, \apj, 462, 316

\bibitem[{{Huggins} \& {Manley}(2005)}]{HugginsM2005}
{Huggins}, P.~J. \& {Manley}, S.~P. 2005, \pasp, 117, 665

\bibitem[{{Jordan} {et~al.}(2005){Jordan}, {Werner}, \&
  {O'Toole}}]{JordanWO2005}
{Jordan}, S., {Werner}, K., \& {O'Toole}, S.~J. 2005, \aap, 432, 273

\bibitem[{{Kerber} {et~al.}(2003){Kerber}, {Mignani}, {Guglielmetti}, \&
  {Wicenec}}]{Kerber+2003}
{Kerber}, F., {Mignani}, R.~P., {Guglielmetti}, F., \& {Wicenec}, A. 2003,
  \aap, 408, 1029

\bibitem[{{Kothes} \& {Landecker}(2004)}]{KothesL2004}
{Kothes}, R. \& {Landecker}, T.~L. 2004, in The Magnetized Interstellar Medium,
  ed. B.~{Uyan{\i}ker}, W.~{Reich}, \& R.~{Wielebinski}, 33--38

\bibitem[{{Landecker} {et~al.}(2000){Landecker}, {Dewdney}, {Burgess}, {Gray},
  {Higgs}, {Hoffmann}, {Hovey}, {Karpa}, {Lacey}, {Prowse}, {Purton}, {Roger},
  {Willis}, {Wyslouzil}, {Routledge}, \& {Vaneldik}}]{Landecker+2000}
{Landecker}, T.~L., {Dewdney}, P.~E., {Burgess}, T.~A., {Gray}, A.~D., {Higgs},
  L.~A., {Hoffmann}, A.~P., {Hovey}, G.~J., {Karpa}, D.~R., {Lacey}, J.~D.,
  {Prowse}, N., {Purton}, C.~R., {Roger}, R.~S., {Willis}, A.~G., {Wyslouzil},
  W., {Routledge}, D., \& {Vaneldik}, J.~F. 2000, \aaps, 145, 509

\bibitem[{{Landecker} {et~al.}(2008)}]{Landecker+2008}
{Landecker}, T.~L. {et~al.} 2008, \aj, in preparation

\bibitem[{{Liebert}(1988)}]{Liebert1998}
{Liebert}, J. 1988, \pasp, 100, 1302

\bibitem[{{Liebert} {et~al.}(2003){Liebert}, {Bergeron}, \&
  {Holberg}}]{Liebert+2003}
{Liebert}, J., {Bergeron}, P., \& {Holberg}, J.~B. 2003, \aj, 125, 348

\bibitem[{{Mathewson} \& {Ford}(1970)}]{MathewsonF1970}
{Mathewson}, D.~S. \& {Ford}, V.~L. 1970, \memras, 74, 139

\bibitem[{{Ng} {et~al.}(2005){Ng}, {Landecker}, {Cazzolato}, {Routledge},
  {Gray}, \& {Reid}}]{Ng+2005}
{Ng}, T., {Landecker}, T.~L., {Cazzolato}, F., {Routledge}, D., {Gray}, A.~D.,
  \& {Reid}, R.~I. 2005, Radio Science, 40, 5014

\bibitem[{{Ohno} \& {Shibata}(1993)}]{OhnoS1993}
{Ohno}, H. \& {Shibata}, S. 1993, \mnras, 262, 953

\bibitem[{{Rand} \& {Kulkarni}(1989)}]{RandK1989}
{Rand}, R.~J. \& {Kulkarni}, S.~R. 1989, \apj, 343, 760

\bibitem[{{Rand} \& {Lyne}(1994)}]{RandL1994}
{Rand}, R.~J. \& {Lyne}, A.~G. 1994, \mnras, 268, 497

\bibitem[{{Rauch} {et~al.}(2007){Rauch}, {Ziegler}, {Werner}, {Kruk},
  {Oliveira}, {Vande Putte}, {Mignani}, \& {Kerber}}]{Rauch+2007}
{Rauch}, T., {Ziegler}, M., {Werner}, K., {Kruk}, J.~W., {Oliveira}, C.~M.,
  {Vande Putte}, D., {Mignani}, R.~P., \& {Kerber}, F. 2007, \aap, 470, 317

\bibitem[{{Reich} {et~al.}(1997){Reich}, {Reich}, \& {F{\"u}rst}}]{ReichRF1997}
{Reich}, P., {Reich}, W., \& {F{\"u}rst}, E. 1997, \aaps, 126, 413

\bibitem[{{Reid} {et~al.}(2008){Reid}, {Gray}, {Landecker}, \&
  {Willis}}]{Reid+2008}
{Reid}, R.~I., {Gray}, A.~D., {Landecker}, T.~L., \& {Willis}, A.~G. 2008,
  Radio Science, 43, 2008

\bibitem[{{Reynolds}(1985)}]{Reynolds1985}
{Reynolds}, R.~J. 1985, \apj, 288, 622

\bibitem[{{Schnitzeler} {et~al.}(2007){Schnitzeler}, {Katgert}, {Haverkorn}, \&
  {de Bruyn}}]{Schnitzeler+2007}
{Schnitzeler}, D.~H.~F.~M., {Katgert}, P., {Haverkorn}, M., \& {de Bruyn},
  A.~G. 2007, \aap, 461, 963

\bibitem[{{Soker}(2002)}]{Soker2002}
{Soker}, N. 2002, \mnras, 336, 826

\bibitem[{{Soker} \& {Dgani}(1997)}]{SokerD1997}
{Soker}, N. \& {Dgani}, R. 1997, \apj, 484, 277

\bibitem[{{Soker} \& {Kastner}(2003)}]{SokerK2003}
{Soker}, N. \& {Kastner}, J.~H. 2003, \apj, 592, 498

\bibitem[{{Soker} \& {Zucker}(1997)}]{SokerZ1997}
{Soker}, N. \& {Zucker}, D.~B. 1997, \mnras, 289, 665

\bibitem[{{Spoelstra}(1984)}]{Spoelstra1984}
{Spoelstra}, T.~A.~T. 1984, \aap, 135, 238

\bibitem[{{Taylor} {et~al.}(2003){Taylor}, {Gibson}, {Peracaula}, {Martin},
  {Landecker}, {Brunt}, {Dewdney}, {Dougherty}, {Gray}, {Higgs}, {Kerton},
  {Knee}, {Kothes}, {Purton}, {Uyan{\i}ker}, {Wallace}, {Willis}, \&
  {Durand}}]{Taylor+2003}
{Taylor}, A.~R., {Gibson}, S.~J., {Peracaula}, M., {Martin}, P.~G.,
  {Landecker}, T.~L., {Brunt}, C.~M., {Dewdney}, P.~E., {Dougherty}, S.~M.,
  {Gray}, A.~D., {Higgs}, L.~A., {Kerton}, C.~R., {Knee}, L.~B.~G., {Kothes},
  R., {Purton}, C.~R., {Uyan{\i}ker}, B., {Wallace}, B.~J., {Willis}, A.~G., \&
  {Durand}, D. 2003, \aj, 125, 3145

\bibitem[{{Tweedy} \& {Kwitter}(1996)}]{TweedyK1996}
{Tweedy}, R.~W. \& {Kwitter}, K.~B. 1996, \apjs, 107, 255

\bibitem[{{Tweedy} {et~al.}(1995){Tweedy}, {Martos}, \&
  {Noriega-Crespo}}]{TweedyMN1995}
{Tweedy}, R.~W., {Martos}, M.~A., \& {Noriega-Crespo}, A. 1995, \apj, 447, 257

\bibitem[{{Uyan{\i}ker}(2004)}]{Uyaniker2004}
{Uyan{\i}ker}, B. 2004, in The Magnetized Interstellar Medium, ed.
  B.~{Uyan{\i}ker}, W.~{Reich}, \& R.~{Wielebinski}, 71--80

\bibitem[{{Uyan{\i}ker} \& {Landecker}(2002)}]{UyanikerL2002b}
{Uyan{\i}ker}, B. \& {Landecker}, T.~L. 2002, \apj, 575, 225

\bibitem[{{Uyan{\i}ker} {et~al.}(2003){Uyan{\i}ker}, {Landecker}, {Gray}, \&
  {Kothes}}]{Uyaniker+2003}
{Uyan{\i}ker}, B., {Landecker}, T.~L., {Gray}, A.~D., \& {Kothes}, R. 2003,
  \apj, 585, 785

\bibitem[{{Vlemmings} {et~al.}(2006){Vlemmings}, {Diamond}, \&
  {Imai}}]{VlemmingsDI2006}
{Vlemmings}, W.~H.~T., {Diamond}, P.~J., \& {Imai}, H. 2006, \nat, 440, 58

\bibitem[{{Vlemmings} {et~al.}(2002){Vlemmings}, {Diamond}, \& {van
  Langevelde}}]{VlemmingsDL2002}
{Vlemmings}, W.~H.~T., {Diamond}, P.~J., \& {van Langevelde}, H.~J. 2002, \aap,
  394, 589

\bibitem[{{Vlemmings} {et~al.}(2005){Vlemmings}, {van Langevelde}, \&
  {Diamond}}]{VlemmingsLD2005}
{Vlemmings}, W.~H.~T., {van Langevelde}, H.~J., \& {Diamond}, P.~J. 2005, \aap,
  434, 1029

\bibitem[{{Wareing} {et~al.}(2007){Wareing}, {Zijlstra}, \&
  {O'Brien}}]{WareingZO2007}
{Wareing}, C.~J., {Zijlstra}, A.~A., \& {O'Brien}, T.~J. 2007, \mnras, 382,
  1233

\bibitem[{{Wieringa} {et~al.}(1993){Wieringa}, {de Bruyn}, {Jansen}, {Brouw},
  \& {Katgert}}]{Wieringa+1993}
{Wieringa}, M.~H., {de Bruyn}, A.~G., {Jansen}, D., {Brouw}, W.~N., \&
  {Katgert}, P. 1993, \aap, 268, 215

\bibitem[{{Willis}(1999)}]{Willis1999}
{Willis}, A.~G. 1999, \aaps, 136, 603

\bibitem[{{Wolleben} {et~al.}(2006){Wolleben}, {Landecker}, {Reich}, \&
  {Wielebinski}}]{Wolleben+2006}
{Wolleben}, M., {Landecker}, T.~L., {Reich}, W., \& {Wielebinski}, R. 2006,
  \aap, 448, 411

\bibitem[{{Wolleben} \& {Reich}(2004)}]{WollebenR2004}
{Wolleben}, M. \& {Reich}, W. 2004, \aap, 427, 537

\end{thebibliography}

% ** Tables here **

\begin{deluxetable}{c c c c c c c}
\tabletypesize{\small}
%\tabletypesize{\scriptsize}
\tablecaption{Polarization Angles in the ``Knots'' of the North-East Arc \label{knotangles}}
\tablewidth{0pt}
\tablehead{
  \colhead{Knot} &
  \colhead{$P.A.$} &
  \colhead{$\sigma_{P.A.}$} &
  \colhead{Extrema} &
  \colhead{$n$} &
  \colhead{$P.I.$} &
  \colhead{$\sigma_{P.I.}$} \\
  \colhead{} &
  \colhead{($\arcdeg$)} &
  \colhead{($\arcdeg$)} &
  \colhead{($\arcdeg$)} &
  \colhead{} &
  \colhead{(K)} &
  \colhead{(K)} \\
  \colhead{} &
  \colhead{[1]} &
  \colhead{[2]} &
  \colhead{[3]} &
  \colhead{[4]} &
  \colhead{[5]} &
  \colhead{[6]}
}
\startdata
1 & $+33.6$ & 13.4    & $+59.7$, $+21.0$ & 10    & 0.139 & 0.017 \\
2 & $+83.7$ & 13.1    & $-74.4$, $+66.0$ & \phn7 & 0.125 & 0.025 \\
3 & $-88.1$ & \phn4.2 & $-83.3$, $+86.8$ & 10    & 0.122 & 0.013 \\
4 & $+52.7$ & \phn2.5 & $+55.6$, $+48.1$ & \phn8 & 0.229 & 0.032 \\
5 & $-79.9$ & \phn2.4 & $-76.8$, $-84.3$ & 10    & 0.175 & 0.015 \\
6 & $-76.6$ & \phn4.7 & $-70.2$, $-82.5$ & 10    & 0.217 & 0.009 \\
7 & $+89.9$ & \phn3.9 & $-83.3$, $+84.9$ & 10    & 0.187 & 0.016 \\
8 & $+65.7$ & \phn2.2 & $+69.7$, $+62.0$ & 10    & 0.220 & 0.013 \\
\enddata
\tablecomments{[1] Mean polarization angle over $n$ pixels; [2]
  Standard deviation in polarization angle over $n$ pixels; [3] Low,
  high value of polarization angle (running clockwise from $P.A.\ =
  0$); [4] Number of pixels used to estimate the mean polarization
  angle; [5] Polarized intensity at center of knot; [6] Standard
  deviation in polarized intensity over $n$ pixels.  Polarization
  angles are modulo 180\arcdeg\@; i.e., angles of $-90$\arcdeg\@ and
  $+90$\arcdeg\@ are equivalent.}
\end{deluxetable}

\clearpage

% ** Figures here **

\end{document}